\documentclass[conference]{IEEEtran}
\IEEEoverridecommandlockouts
\usepackage{cite}
\usepackage{amsmath,amssymb,amsfonts}
\usepackage{mathtools}
\usepackage{graphicx}
\usepackage{textcomp}
\usepackage{xcolor}
\usepackage{mathcomSTEv4}
\usepackage{mathtools}

\usepackage{array}
\usepackage{color}
\usepackage{bm}
\usepackage{mathrsfs}
\usepackage{amsmath,amsfonts,bm,amsfonts}
\usepackage{mathrsfs}
\usepackage{algorithmic}
\usepackage{algorithm}
\usepackage{epsf}
\usepackage{xpatch}
\usepackage{setspace} 

\usepackage{caption}
\usepackage{subcaption}
\usepackage{enumitem,kantlipsum}
\usepackage{multirow,booktabs}
\usepackage{latexsym}
\usepackage{hyperref}
\usepackage{cleveref}
\usepackage[export]{adjustbox}
\usepackage{dsfont}
\def\BibTeX{{\rm B\kern-.05em{\sc i\kern-.025em b}\kern-.08em
    T\kern-.1667em\lower.7ex\hbox{E}\kern-.125emX}}
\begin{document}

\title{HiKO: A Hierarchical Framework for Beyond-Second-Order KO Codes}

\author{
	\IEEEauthorblockN{Shubham Srivastava, and Adrish Banerjee} %\thanks{The work is accepted in IEEE International Symposium on Information Theory (ISIT) 2023.} 
	\IEEEauthorblockA{ Department of Electrical Engineering, Indian Institute of Technology Kanpur, India \\
		Email: \{shubhsr, adrish\}@iitk.ac.in           
	}
}

\maketitle

\begin{abstract}
	This paper introduces HiKO (Hierarchical Kronecker Operation), a novel framework for training high-rate neural error-correcting codes that enables KO codes to outperform Reed-Muller codes beyond second order.  To our knowledge, this is the first attempt to extend KO codes beyond second order. While conventional KO codes show promising results for low-rate regimes (r$<$2), they degrade at higher rates—a critical limitation for practical deployment. Our framework incorporates three key innovations: (1) a hierarchical training methodology that decomposes complex high-rate codes into simpler constituent codes for efficient knowledge transfer, (2) enhanced neural architectures with dropout regularization and learnable skip connections tailored for the Plotkin structure, and (3) a progressive unfreezing strategy that systematically transitions from pre-trained components to fully optimized integrated codes. Our experiments show that HiKO codes consistently outperform traditional Reed-Muller codes across various configurations, achieving notable performance improvements for third-order (r=3) and fourth-order (r=4) codes. Analysis reveals that HiKO codes successfully approximate Shannon-optimal Gaussian codebooks while preserving efficient decoding properties. This represents the first successful extension of KO codes beyond second-order, opening new possibilities for neural code deployment in high-throughput communication systems.
\end{abstract}


\section{Introduction}
Channel coding has been foundational in the development of reliable communications systems, with landmark codes like Reed-Muller (RM) \cite{1057465}, BCH\cite{bose1960class}, Turbo \cite{397441}, LDPC \cite{gallager1963low}, and Polar codes \cite{4595172} each representing significant mathematical breakthroughs. These codes approach theoretical performance limits through diverse mathematical structures.

\subsection{Machine Learning for Channel Coding}
Recent advances in machine learning have opened new avenues for innovation in channel coding \cite{Hebbar_2022, 9174106, 9406115, 9281328}. Initial efforts focused on enhancing traditional decoding algorithms, with Nachmani et al. pioneering neural belief propagation decoders \cite{7852251} and subsequent works extending to various code families \cite{8242643, 8815400, 8445986, Hebbar_2022}. 
End-to-end learned codes emerged through channel autoencoders \cite{7886039, 8054694}, with significant advancements including Turbo Autoencoders \cite{8890904}, DRF codes \cite{mashhadi2021drf}, and adaptive approaches like MIND \cite{8815537}, TurboNet \cite{9145754}, and Li et al.'s residual learning framework \cite{9109744}.
A breakthrough came with KO (Kronecker Operation) codes \cite{pmlr-v139-makkuva21a}, which structured neural architectures based on classical code families. By using the computation tree of Reed-Muller codes as a skeleton, KO codes successfully generalized the underlying Kronecker operation, enabling powerful nonlinear codes that outperformed classical counterparts at short-to-medium block lengths. Further developments include Dense KO codes \cite{10619329}, ProductAE \cite{vahid2021productae}, Jamali et al.'s subcodes \cite{jamali2021reed}, and DeepPolar codes \cite{hebbar2024deeppolar}.

\subsection{Challenges at Higher Code Rates}
KO codes work much better than RM codes at low rates (small $r$), but their benefits decrease as the code rate goes up (large $r$). At rates $r>2$, KO codes often do not outperform RM codes and can even do worse.

This happens due to several reasons:
\begin{itemize}
	\item \textbf{Harder optimization}: More parameters make it easier to get stuck in bad  local minima.
	\item \textbf{Unstable training}: Deeper networks needed for high rates can have vanishing or exploding gradients.
	\item \textbf{Limited architecture}: Standard KO designs cannot capture the complex patterns in high-rate codes.
	\item \textbf{High computation}: Training large codes needs a lot of computing power, so fewer experiments are possible.
\end{itemize}
These issues lead us to use a hierarchical training method that uses the natural structure of RM and KO codes.

\subsection{Our Contributions}
This paper makes several significant contributions:
\begin{itemize}
	\item \textbf{Hierarchical Framework:} We introduce HiKO, enabling efficient knowledge transfer from simpler constituent codes to complex higher-rate structures.
	\item \textbf{Enhanced Architecture:} We design specialized neural networks with learnable skip connections and dropout regularization for higher-order codes.
	\item \textbf{Progressive Training:} We implement a three-phase protocol with progressive unfreezing and adaptive learning rates.
	\item \textbf{Performance:} Our extensive testing of HiKO codes demonstrates better BER performance across diverse configurations(KO(8,3), KO(8,4), KO(9,3), and KO(9,4)), marking the first successful implementation of KO codes that achieve excellence at higher rates (r$>$2).
\end{itemize}

The remainder of this paper is organized as follows: Section~\ref{sec:background} provides essential background on Reed-Muller codes, Kronecker Operation codes, and decoding algorithms. Section~\ref{sec:hiko} introduces our hierarchical framework. Section~\ref{subsec:ber_analysis} presents experimental results and analysis, while Section~\ref{sec:conclusion} discusses conclusions and future research directions.

\subsection{Reproducibility}
For reproducibility, the code is shared at \url{https://github.com/shubhamsrivast4u/HiKO-Codes}

\section{Background} \label{sec:background}

This section provides the necessary foundation for understanding our contribution by reviewing Reed-Muller codes, Kronecker Operation (KO) codes, and their respective decoding algorithms.

\subsection{Reed-Muller Codes}

Reed-Muller (RM) codes are a family of linear error-correcting codes parameterized by a variable size $m \in \mathbb{Z}^+$ and an order $r \in \mathbb{Z}^+$ with $r \leq m$, denoted as RM($m$, $r$). The code maps binary information bits $\mathbf{m} \in \{0,1\}^k$ to codewords $\mathbf{x} \in \{0,1\}^n$, where the code length $n = 2^m$ and the number of information bits $k = \sum_{i=0}^{r} \binom{m}{i}$.

RM codes are defined through the recursive application of the Plotkin construction \cite{1057584}. The basic mapping $\text{Plotkin}: \{0,1\}^\ell \times \{0,1\}^\ell \to \{0,1\}^{2\ell}$ is defined as:

\begin{equation}
	\text{Plotkin}(\mathbf{u}, \mathbf{v}) = (\mathbf{u}, \mathbf{u} \oplus \mathbf{v})
\end{equation}

with $\oplus$ representing coordinate-wise XOR and $(\cdot,\cdot)$ denoting concatenation. RM codes are recursively defined as:
\begin{equation}
	\text{RM}(m, r) = \{(\mathbf{u}, \mathbf{u} \oplus \mathbf{v})\},
\end{equation}
where $\mathbf{u} \in \text{RM}(m-1, r)$ and $\mathbf{v} \in \text{RM}(m-1, r-1)$.
The RM($m$, 0) is a repetition code that repeats a single information bit $2^m$ times.

\subsection{Dumer's Recursive Decoder for RM Codes}

Dumer's algorithm \cite{1291729} efficiently decodes RM codes by exploiting the Plotkin structure. For a received noisy codeword $\mathbf{y} \in \mathbb{R}^{2^m}$, the decoder computes log-likelihood ratios (LLRs):
\begin{equation}
	L_i = \log \frac{P[\mathbf{y}_i|\mathbf{x}_i=0]}{P[\mathbf{y}_i|\mathbf{x}_i=1]}, \quad i = 1, \ldots, 2^m
\end{equation}

To decode an RM($m$, $r$) codeword $\mathbf{x} = (\mathbf{u}, \mathbf{u} \oplus \mathbf{v})$, the algorithm first obtains soft information for the left sub-codeword $\mathbf{v} \in \text{RM}(m-1, r-1)$ using:
\begin{equation}
	\mathbf{L}_v = \text{LSE}(\mathbf{L}_{1:2^{m-1}}, \mathbf{L}_{2^{m-1}+1:2^m})
\end{equation}
where $\text{LSE}(a, b) = \log\left(\frac{1+e^{a+b}}{e^a+e^b}\right)$ for $a, b \in \mathbb{R}$.

After decoding $\mathbf{v}$ to obtain $\hat{\mathbf{v}}$, the algorithm computes the soft information for $\mathbf{u} \in \text{RM}(m-1, r)$ by:
\begin{equation}
	\mathbf{L}_u = \mathbf{L}_{1:2^{m-1}} \oplus_{\hat{\mathbf{v}}} \mathbf{L}_{2^{m-1}+1:2^m} = \mathbf{L}_{1:2^{m-1}} + (-1)^{\hat{\mathbf{v}}} \mathbf{L}_{2^{m-1}+1:2^m}
\end{equation}

The algorithm continues recursively until reaching leaf nodes, where it performs MAP decoding, enabling efficient decoding with complexity $O(n \log n)$.

\subsection{KO Codes}

KO codes \cite{pmlr-v139-makkuva21a} build upon RM codes by generalizing the Plotkin construction through neural networks. A KO code is denoted as KO($m$, $r$, $\mathbf{g}_\theta$, $\mathbf{f}_\phi$), parameterized by integers $m, r$, the KO encoder $\mathbf{g}_\theta$, and the KO decoder $\mathbf{f}_\phi$. The code dimension $k = \sum_{i=0}^{r} \binom{m}{i}$ is identical to RM codes.

\subsubsection{KO Encoder}

The KO encoder inherits the Plotkin tree structure of the RM encoder. For each internal node $i$, the neural network $g_i$ applies a coordinate-wise nonlinear transformation mapping $(\mathbf{u}, \mathbf{v}) \to g_i(\mathbf{u}, \mathbf{v}) \in \mathbb{R}^{2\ell}$, where:
\begin{equation}
	g_i(\mathbf{u}, \mathbf{v}) = (\mathbf{u}, \tilde{g}_i(\mathbf{u}, \mathbf{v}) + \mathbf{u} \oplus \mathbf{v})
\end{equation}

Here, $\tilde{g}_i: \mathbb{R}^2 \to \mathbb{R}$ is a neural network applied coordinate-wise, generalizing the classical Plotkin mapping.

\subsubsection{KO Decoder}

The KO decoder matches the encoder's structure while generalizing Dumer's algorithm. It consists of neural networks $(f_{2i-1}, f_{2i})$ for each node $i$ in the Plotkin tree. Given received vector $\mathbf{y} = (\mathbf{y}_1, \mathbf{y}_2) \in \mathbb{R}^{2\ell}$, the decoder computes:

\begin{align}
	f_{2i-1}(\mathbf{y}_1, \mathbf{y}_2) &= \tilde{f}_{2i-1}(\mathbf{y}_1, \mathbf{y}_2) + \text{LSE}(\mathbf{y}_1, \mathbf{y}_2) \\
	f_{2i}(\mathbf{y}_1, \mathbf{y}_2, \mathbf{y}_v, \hat{\mathbf{v}}) &= \tilde{f}_{2i}(\mathbf{y}_1, \mathbf{y}_2, \mathbf{y}_v, \hat{\mathbf{v}}) + \mathbf{y}_1 + (-1)^{\hat{\mathbf{v}}} \mathbf{y}_2
\end{align}

where $\mathbf{y}_v = f_{2i-1}(\mathbf{y}_1, \mathbf{y}_2)$ is soft information for the left branch, $\hat{\mathbf{v}}$ is the decoded estimate of the left branch codeword, and $\tilde{f}_{2i-1}, \tilde{f}_{2i}$ are neural networks applied coordinate-wise.
The decoder processes the Plotkin tree depth-first, first decoding the leftmost nodes then proceeding right. At leaf nodes, it employs soft-MAP decoding to enable end-to-end training.

\section{HiKO: Hierarchical Training for High-Rate KO Codes}
\label{sec:hiko}
We introduce HiKO codes, a framework for learning high-rate Kronecker Operation neural codes through hierarchical training. Our approach leverages the recursive structure of RM and KO codes to enable compositional learning, where knowledge from simpler, lower-dimensional codes can be transferred to more complex, higher-dimensional codes.

\subsection{Initial Approaches to Higher-Rate Extensions}
\label{subsec:initial_approaches}

Before developing our hierarchical methodology, we explored two direct extension approaches that provided valuable insights despite their limitations.
\subsubsection{Generalized LLR Propagation Framework}
\label{subsubsec:llr_propagation}
Our first approach extended the decoder through a generalized LLR propagation framework, leveraging the dual structure of Plotkin construction for soft-decoding. We developed a recursive LLR transformation function $\mathcal{M}_r$ that maps received LLRs to information bit LLRs with enhanced numerical stability:
\begin{align}
	\mathcal{M}_r(\mathbf{L}, m) = 
	\begin{cases}
		&\mathbf{L}.\text{repeat}(1, 2^m),  r = 0 \\
		&\mathcal{S}(\mathcal{C}(\mathbf{L}, G_m)),  r = 1 \\
		&\mathcal{F}(\mathbf{L}_u, \mathbf{L}_v, r, m),  \text{otherwise}
	\end{cases}
\end{align}
where $\mathcal{S}$ applies sign processing with numerical stability constraints, $\mathcal{C}$ represents tensor contraction with generator matrix $G_m$, and $\mathcal{F}$ combines LLRs from recursive calls using a tanh-log formulation to avoid numerical instability:
\begin{align}
	\mathcal{F}(\mathbf{L}_u, \mathbf{L}_v, r, m) = \mathcal{B}(\mathcal{M}_r(\mathbf{L}_u, m-1), \mathcal{M}_{r-1}(\mathbf{L}_v, m-1)) \nonumber
\end{align}
Here, $\mathcal{B}$ represents our stabilized combination function that applies appropriate scaling and clamping to prevent gradient explosion.

\begin{figure}[t]
	\centering
	\includegraphics[width=0.9\columnwidth]{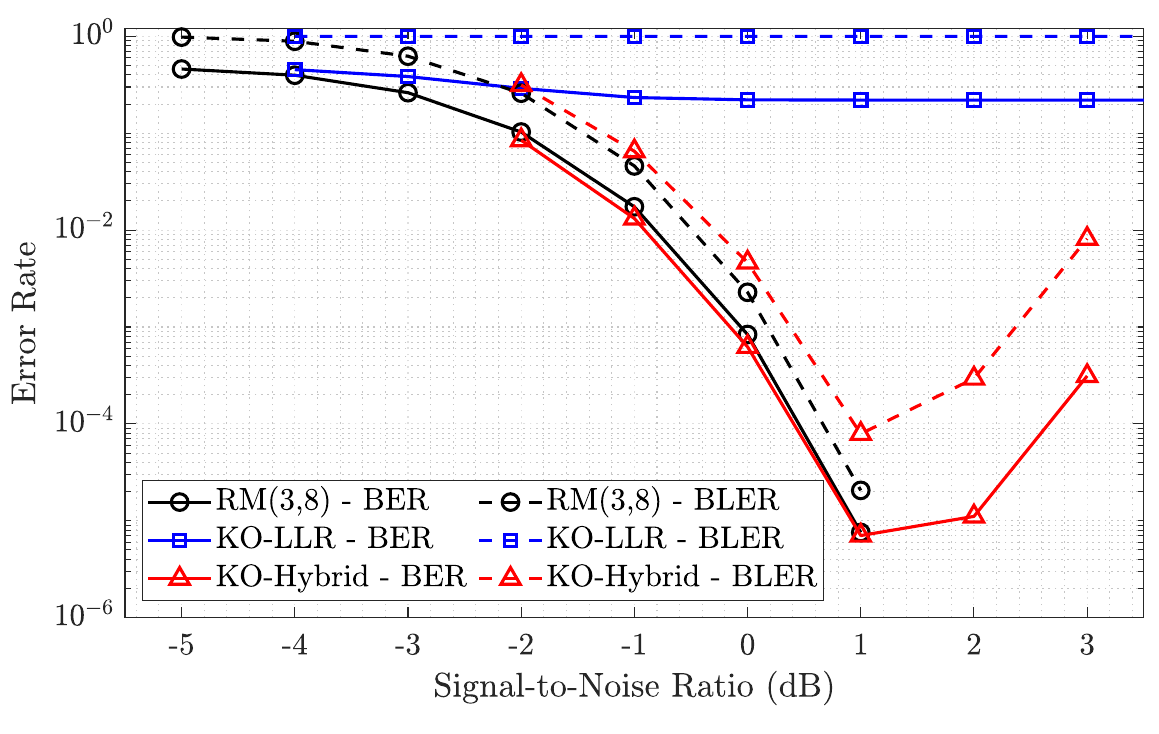}
	\caption{Performance comparison of classical Reed-Muller RM(3,8) code and two initial variants of higher-rate Kronecker Operation (KO) codes. KO-LLR utilizes generalized LLR propagation while KO-Hybrid implements a hybrid recursive decoding structure.}
	\label{fig:higher_rate_ko}
\end{figure}
Figure~\ref{fig:higher_rate_ko} shows that the BER values in KO-LLR variant stabilizes around 0.22 across the tested SNR range for KO(3,8) code. While this represents an improvement over random guessing (BER = 0.5), it still underperforms compared to classical RM codes. This suggests that recursive structure accumulates error with depth, motivating our development of alternative approaches that better maintain gradient flow.
\subsubsection{Hybrid Recursive Decoding Approach}
\label{subsubsec:hybrid_recursive}
Our second approach implemented a hybrid recursive decoding structure combining soft and hard decisions. Instead of propagating LLRs directly, we first decoded the left branch, converted to hard decisions, then used these for decoding the right branch:
\begin{equation}
	\text{RecDec}(\mathbf{L}, r, m, \mathbf{f}_{\phi}) = \text{Combine}(\hat{\mathbf{m}}_{u}, \hat{\mathbf{m}}_{v}) 
\end{equation} 
\begin{align}
	\text{where: } 
	\mathbf{L}_{v} &= f_{v}(\mathbf{L}_{\text{left}}, \mathbf{L}_{\text{right}}) + \text{LSE}(\mathbf{L}_{\text{left}}, \mathbf{L}_{\text{right}}) \nonumber \\
	\hat{\mathbf{m}}_{v} &= \text{RecDec}(\mathbf{L}_{v}, r-1, m-1, \mathbf{f}_{\phi}) \nonumber \\
	\hat{\mathbf{v}} &= \mathcal{E}(\hat{\mathbf{m}}_{v}.\text{sign}(), r-1, m-1) \nonumber \\
	\mathbf{L}_{u} &= f_{u}(\mathbf{L}_{\text{left}}, \mathbf{L}_{\text{right}}, \mathbf{L}_{v}, \hat{\mathbf{v}}) + \mathbf{L}_{\text{left}} + \hat{\mathbf{v}} \odot \mathbf{L}_{\text{right}} \nonumber \\
	\hat{\mathbf{m}}_{u} &= \text{RecDec}(\mathbf{L}_{u}, r, m-1, \mathbf{f}_{\phi}) 
\end{align}
where $\mathcal{E}(\cdot)$ represents the classical Plotkin encoding applied to hard-decoded bits.
As shown in Figure~\ref{fig:higher_rate_ko}, this hybrid approach demonstrated dramatically improved training stability. At moderate to high SNR values (0-3 dB), the KO-Hybrid achieves BER performance that matches or slightly exceeds the classical RM(3,8) code. For instance, at 0 dB, the KO-Hybrid achieves a BER of $6.24 \times 10^{-4}$ compared to $8.37 \times 10^{-4}$ for the RM code.
However, these improvements are inconsistent across the operating range. Our empirical results highlight the significant challenges in training high-rate neural codes: unstable gradient flow, complex optimization landscapes, and the difficulty of maintaining performance across wide operating ranges. These observations provide strong motivation for our development of a hierarchical training methodology.

\begin{figure*}[htbp] 
	\centering 
	\begin{subfigure}[b]{0.2\textwidth} 
		\centering 
		\includegraphics[width=\textwidth]{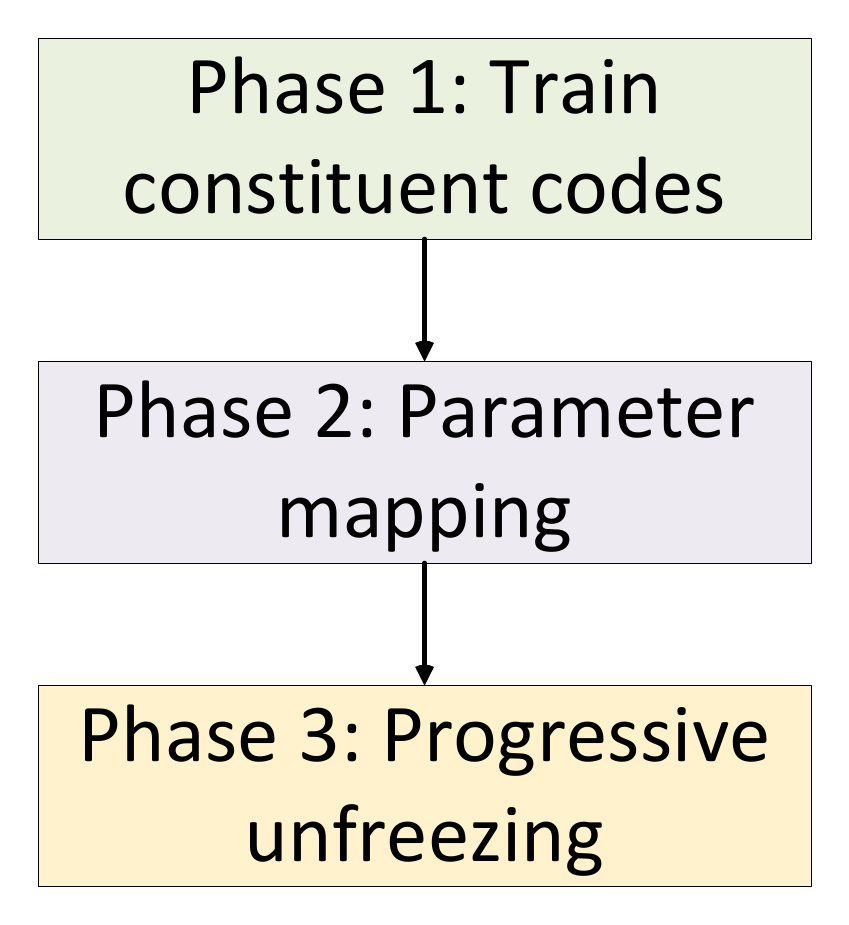} 
		\caption{} 
		\label{fig:phasetraining} 
	\end{subfigure} 
	\hfill 
	\begin{subfigure}[b]{0.39\textwidth} 
		\centering 
		\includegraphics[width=\textwidth]{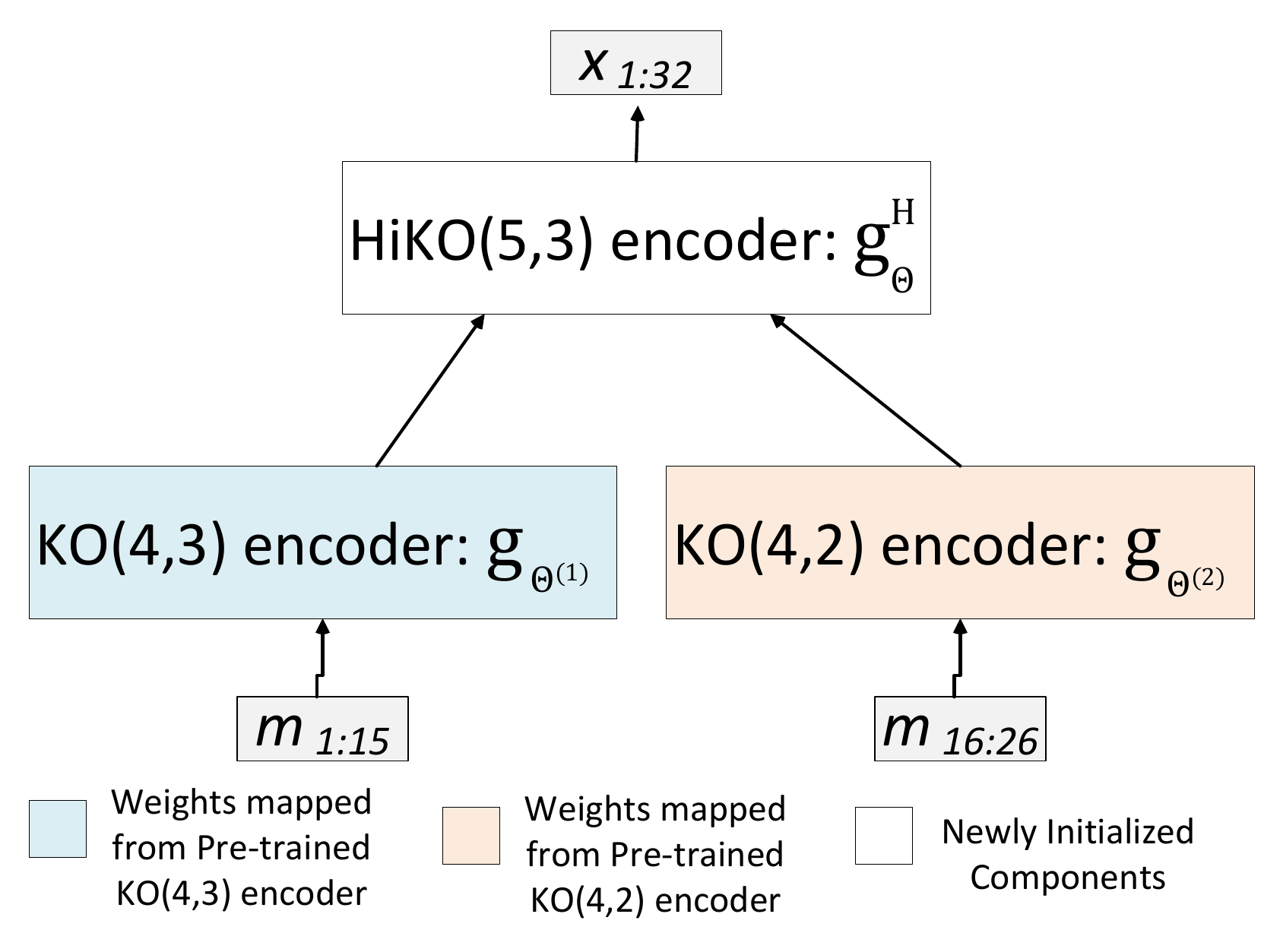} 
		\caption{} 
		\label{fig:encoder} 
	\end{subfigure} 
	\hfill 
	\begin{subfigure}[b]{0.39\textwidth} 
		\centering 
		\includegraphics[width=\textwidth]{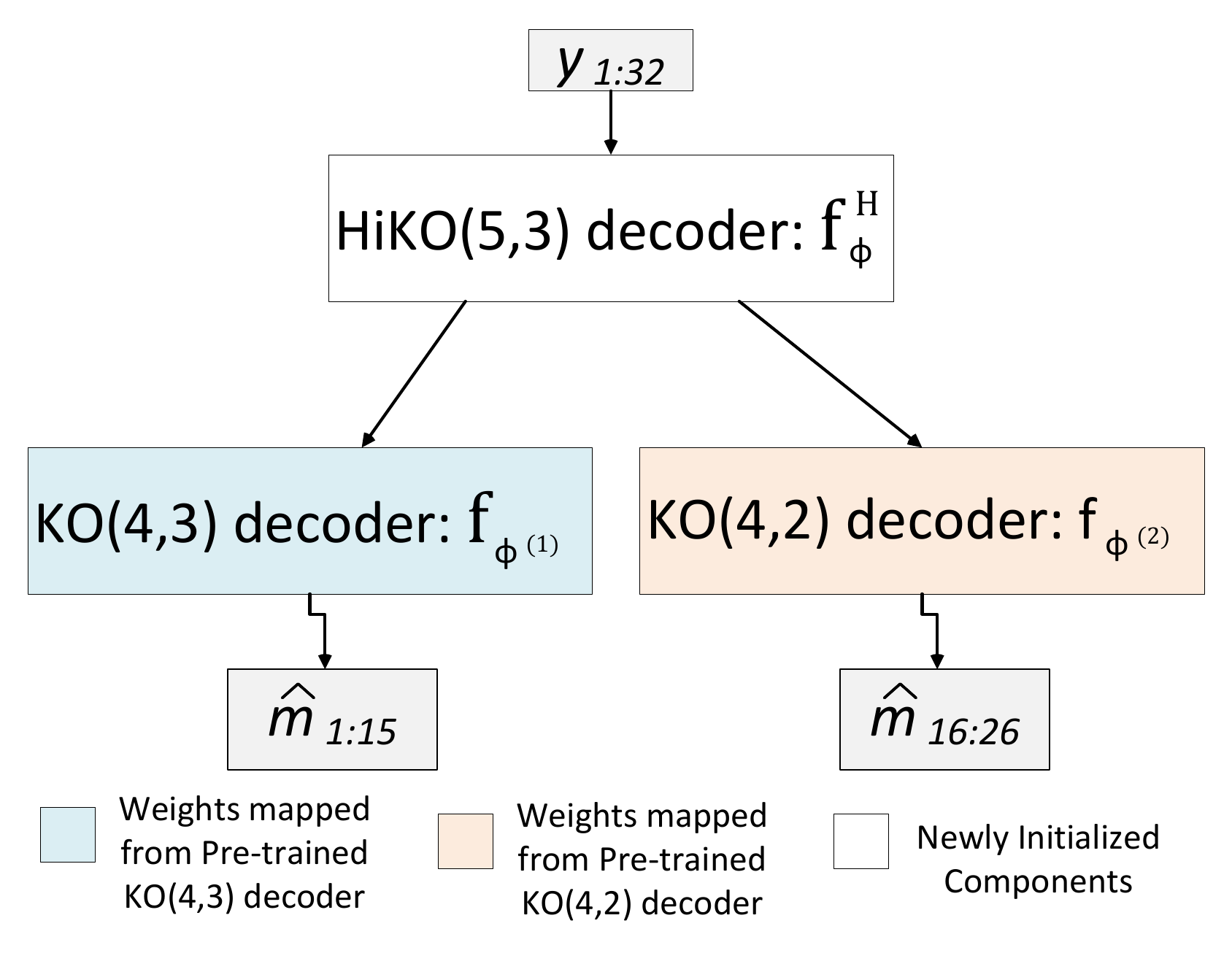} 
		\caption{} 
		\label{fig:decoder} 
	\end{subfigure} 
	\caption{ The HiKO hierarchical training framework: (a) The three-phase training process for HiKO, showing the progression from constituent code training to parameter mapping and finally progressive unfreezing. (b) The encoder architecture for HiKO(5,3), demonstrating how weights from pre-trained KO(4,3) and KO(4,2) constituent encoders are mapped to appropriate positions, with newly initialized components for the remaining structure. (c) The corresponding decoder architecture, showing similar weight mapping patterns from pre-trained constituent decoders to form the hierarchical structure.} 
	\label{fig:hierarchical_ko} 
\end{figure*}

\subsection{Hierarchical Encoder-Decoder Architecture}
Building on the insights from our initial approaches, particularly the Hybrid Recursive Decoding framework which demonstrated improved stability, HiKO adopts this decoding mechanism while enhancing it with hierarchical training and specialized neural architectures. The HiKO framework preserves the sequential soft-to-hard decision flow pattern from the hybrid approach, but incorporates constituent code knowledge transfer and regularization techniques to overcome the training challenges at higher rates.

\label{subsec:hierarchical_architecture}
\subsubsection{Encoder Structure}
\label{subsubsec:encoder_structure}
The HiKO encoder maintains the Plotkin tree structure of standard KO codes with two critical enhancements, as illustrated in Figure~\ref{fig:encoder}:
\begin{enumerate}
	\item \textbf{Parameter sharing across levels}: Neural modules are initialized with pre-trained weights from lower-dimensional constituent codes.
	\item \textbf{Enhanced neural components}: We upgrade standard KO networks with improved dropout regularization and learnable skip connections:
	\begin{equation}
		g_i^{\text{H}}(\mathbf{u}, \mathbf{v}) = (\mathbf{u}, \tilde{g}_i^{\text{H}}(\mathbf{u}, \mathbf{v}) + \alpha_i \cdot (\mathbf{u} \oplus \mathbf{v}))
	\end{equation}
	where $\tilde{g}_i^{\text{H}}$ is a neural network with SELU activations and dropout, and $\alpha_i$ is a learnable scaling parameter.
\end{enumerate}
For a target HiKO$(m,r)$ code with constituent codes $\{\text{KO}(m_j,r_j)\}_{j=1}^{K}$, the encoder mapping is:
\begin{equation}
	g_i^{\text{H}}(\mathbf{u}, \mathbf{v}) = 
	\begin{cases}
		g_{c(i)}^{(j)}(\mathbf{u}, \mathbf{v}) & \text{if } i \in \mathcal{M} \\
		g_i^{\text{H}}(\mathbf{u}, \mathbf{v}) & \text{otherwise}
	\end{cases}
\end{equation}
where:
\begin{itemize}
	\item $g_{c(i)}^{(j)}$ refers to the function from the pre-trained KO$(m_j,r_j)$ code corresponding to node $i$
	\item $c(i)$ is the mapping function that identifies the corresponding index in the constituent code
	\item $\mathcal{M}$ represents the set of node indices in the target code that correspond to pretrained components
\end{itemize}
\subsubsection{Decoder Structure}
\label{subsubsec:decoder_structure}
Similarly, the HiKO decoder follows Dumer's algorithm structure but with parameter sharing from pre-trained constituent decoders, as shown in Figure~\ref{fig:decoder}:
\begin{equation}
	f_l^{\text{H}}(\cdot) = 
	\begin{cases}
		f_{c(l)}^{(j)}(\cdot) & \text{if } l \in \mathcal{M} \\
		f_l^{\text{H}}(\cdot) & \text{otherwise}
	\end{cases}
\end{equation}
where $f_{c(l)}^{(j)}$ refers to the function from the pre-trained KO$(m_j,r_j)$ decoder corresponding to node $l$.

The enhanced decoder neural networks are defined as:
\begin{align}
	f_{2i-1}^{\text{H}}(\mathbf{y}_1, \mathbf{y}_2) &= \tilde{f}_{2i-1}^{\text{H}}(\mathbf{y}_1, \mathbf{y}_2) + \text{LSE}(\mathbf{y}_1, \mathbf{y}_2) \\
	f_{2i}^{\text{H}}(\mathbf{y}_1, \mathbf{y}_2, \mathbf{y}_v, \hat{\mathbf{v}}) &= \tilde{f}_{2i}^{\text{H}}(\mathbf{y}_1, \mathbf{y}_2, \mathbf{y}_v, \hat{\mathbf{v}}) + \mathbf{y}_1 + (-1)^{\hat{\mathbf{v}}} \mathbf{y}_2
\end{align}
where:
\begin{itemize}
	\item $\mathbf{y} = (\mathbf{y}_1, \mathbf{y}_2)$ denotes the received LLR vector split into left and right halves
	\item $\mathbf{y}_v$ is the soft information for the left branch
	\item $\hat{\mathbf{v}}$ is the decoded estimate of the left branch codeword
	\item $\tilde{f}_{2i-1}^{\text{H}}$ and $\tilde{f}_{2i}^{\text{H}}$ are neural networks with dropout regularization, SELU activations, and 120 hidden units.
\end{itemize}

\subsection{Progressive Training Algorithm}
\label{subsec:training_algorithm}
Our hierarchical training procedure consists of three phases as illustrated in Figure~\ref{fig:phasetraining}.The first phase trains constituent codes independently. The second phase maps pre-trained components to appropriate positions in the target code. The final phase employs progressive unfreezing and adaptive learning rates to effectively optimize the integrated model.
\subsection{Parameter Mapping and Progressive Unfreezing}
\label{subsec:parameter_mapping}
A critical aspect of HiKO is correct parameter mapping from constituent codes to the target high-rate code. Given that the Plotkin tree for HiKO$(m,r)$ contains nodes corresponding to smaller KO$(m_j,r_j)$ codes, we systematically:
\begin{enumerate}
	\item Identify nodes in each constituent code's encoder and decoder trees
	\item Locate corresponding subtrees within the target HiKO$(m,r)$ structure
	\item Transfer learned parameters from each constituent node to its corresponding position
\end{enumerate}
For effective adaptation, we employ a progressive unfreezing schedule:
$
	t_j = \left\lfloor \frac{j \cdot T}{K + 1} \right\rfloor
$,
where $T$ is total training epochs, $K$ is the number of constituents, and $t_j$ is the epoch to unfreeze the $j$-th component.
We implement adaptive learning rates based on component status:
\begin{equation}
	\eta_t^{(j)} = 
	\begin{cases}
		0 & \text{if component is frozen} \\
		\mathscr{S}(2 \times 10^{-4}, t) & \text{if newly initialized} \\
		\mathscr{S}(1 \times 10^{-4}, t) & \text{if pre-trained and unfrozen}
	\end{cases}
\end{equation}

where $\mathscr{S}(\eta_{\max}, t)$ represents our cyclical learning rate scheduler with initial value $\eta_{\max}/25$, increasing to $\eta_{\max}$ during the first 30\% of training, then decreasing to $\eta_{\max}/(25 \times 10000)$ by the end.

Through this hierarchical approach, HiKO effectively addresses the challenges of training high-rate KO codes. For our analysis of KO(8,3), KO(8,4), KO(9,3), and KO(9,4) codes, we utilize pretrained KO(4,3) and KO(4,2) constituent codes as building blocks.  These constituent codes were independently trained for 200 epochs before integration into the hierarchical structure. The architecture shown in Figures~\ref{fig:encoder} and \ref{fig:decoder} illustrates the principle that scales to these larger codes. The three-phase training process, shown in   Figures~\ref{fig:encoder},  enables effective knowledge transfer from simpler codes to more complex ones, resulting in significant performance improvements.
\begin{algorithm}[ht!]
	\caption{Hierarchical Training for HiKO Codes}
	\label{alg:hiko_training}
	\begin{algorithmic}[1]
		\REQUIRE Code parameters $(m, r)$, constituent codes $\{(m_j, r_j)\}_{j=1}^{K}$, training SNRs $\sigma_{\text{enc}}^2$ and $\sigma_{\text{dec}}^2$, epochs $T$
		
		\STATE \textbf{Phase 1:} Train constituent codes
		\FOR{$j = 1$ to $K$}
		\STATE Train KO$(m_j, r_j)$ encoder $\mathbf{g}_{\theta^{(j)}}$ and decoder $\mathbf{f}_{\phi^{(j)}}$ independently
		\STATE Save optimized parameters $\theta^{(j)}$ and $\phi^{(j)}$
		\ENDFOR
		
		\STATE \textbf{Phase 2:} Initialize HiKO$(m, r)$ architecture
		\STATE Create encoder $\mathbf{g}_\theta^{\text{H}}$ and decoder $\mathbf{f}_\phi^{\text{H}}$ for HiKO$(m, r)$
		\FOR{$j = 1$ to $K$}
		\STATE Map parameters from $\mathbf{g}_{\theta^{(j)}}$ to corresponding positions in $\mathbf{g}_\theta^{\text{H}}$
		\STATE Map parameters from $\mathbf{f}_{\phi^{(j)}}$ to corresponding positions in $\mathbf{f}_\phi^{\text{H}}$
		\STATE Freeze these transferred parameters initially
		\ENDFOR
		\STATE Initialize remaining parameters randomly
		
		\STATE \textbf{Phase 3:} Train with progressive unfreezing
		\STATE Compute unfreezing schedule $\{t_j\}_{j=1}^{K}$ with $t_j = \lfloor \frac{j \cdot T}{K + 1} \rfloor$
		\FOR{$t = 1$ to $T$}
		\STATE Check schedule and unfreeze components if needed
		\STATE Update learning rates according to adaptive schedule $\eta_t^{(j)}$
		\STATE Train decoder for $T_{\text{dec}}$ iterations using noise $\sigma_{\text{dec}}^2$
		\STATE Train encoder for $T_{\text{enc}}$ iterations using noise $\sigma_{\text{enc}}^2$
		\STATE Validate performance and save best model
		\ENDFOR
		
		\ENSURE Trained HiKO encoder $\mathbf{g}_\theta^{\text{H}}$ and decoder $\mathbf{f}_\phi^{\text{H}}$
		\RETURN Final model parameters $\theta^*$ and $\phi^*$
	\end{algorithmic}
\end{algorithm}

\section{Experimental Results and Analysis}

This section presents experimental results demonstrating the performance of our proposed HiKO codes compared to classical Reed-Muller codes across various parameter configurations.

\subsection{Experimental Setup}
For each code configuration, we utilized the following protocol:
\begin{itemize}
	\item Training data: Randomly generated message bits transmitted over AWGN channels
	\item Training SNR: 0 dB for encoder and -1 dB for decoder
	\item Batch size: 4000 with mini-batches of 1000
	\item Training epochs: 300 with progressive unfreezing
	\item Testing: $10^7$ random message bits per SNR point
\end{itemize}

\subsection{BER Performance Analysis}
\label{subsec:ber_analysis}

\begin{figure}[t]
	\centering
	\includegraphics[width=0.95\columnwidth]{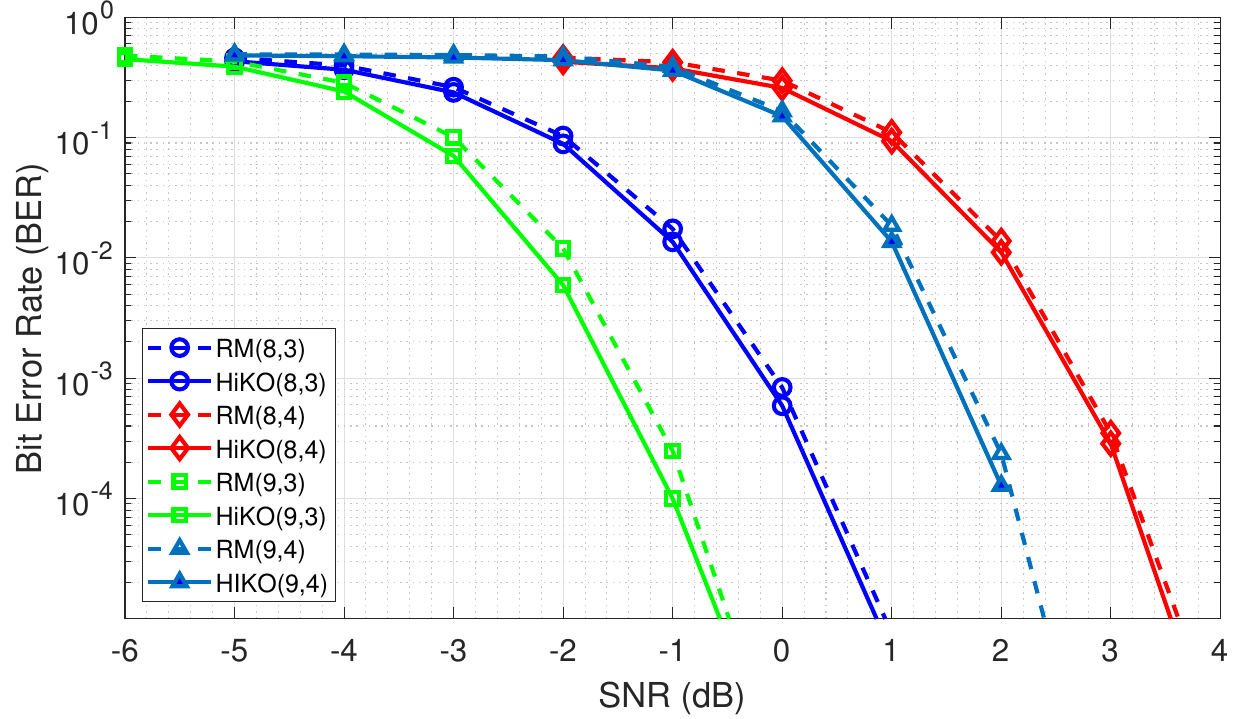}
	\caption{BER performance comparison between classical Reed-Muller codes and hierarchically trained KO codes.}
	\label{fig:ber_comparison}
\end{figure}

Figure \ref{fig:ber_comparison} presents the bit error rate (BER) performance of HiKO codes compared to classical RM codes. For the HiKO(8,3) vs. RM(8,3) comparison, at low SNR (-5 dB), HiKO(8,3) achieves a BER of $4.31 \times 10^{-1}$ compared to $4.60 \times 10^{-1}$ for RM(8,3). This advantage becomes more pronounced as SNR increases, with BER values at 0 dB of $5.90 \times 10^{-4}$ versus $8.37 \times 10^{-4}$. In the HiKO(8,4) vs. RM(8,4) configuration, we observe significant improvement across all measured SNR points. At 1 dB, HiKO(8,4) exhibits a BER of $9.33 \times 10^{-2}$ versus $1.10 \times 10^{-1}$ for RM(8,4), with the performance gap widening at 3 dB to $2.85 \times 10^{-4}$ versus $3.49 \times 10^{-4}$. The most dramatic improvements appear in the HiKO(9,3) vs. RM(9,3) comparison with the larger block length. At -2 dB, HiKO(9,3) achieves a BER of $5.93 \times 10^{-3}$ compared to $1.20 \times 10^{-2}$ for RM(9,3), and at -1 dB, the performance advantage increases further with HiKO(9,3) reaching $9.98 \times 10^{-5}$ versus $2.48 \times 10^{-4}$ for RM(9,3). Similar improvements were observed in HiKO(9,4) configuration.
Our results demonstrate SNR gains ranging from 0.1 to 0.3 dB at a BER of $10^{-4}$. These improvements confirm that our hierarchical training approach successfully overcomes the performance limitations of standard KO codes at rates $r>2$. Comparative analysis reveals that the performance advantages of HiKO codes over RM codes become increasingly pronounced with larger block lengths and higher SNR values, with HiKO(9,3) exhibiting the most substantial improvement at operating points of practical interest.

\subsection{Codeword Distance Distribution Analysis}

\begin{figure}[t]
	\centering
	\includegraphics[width=0.85\columnwidth]{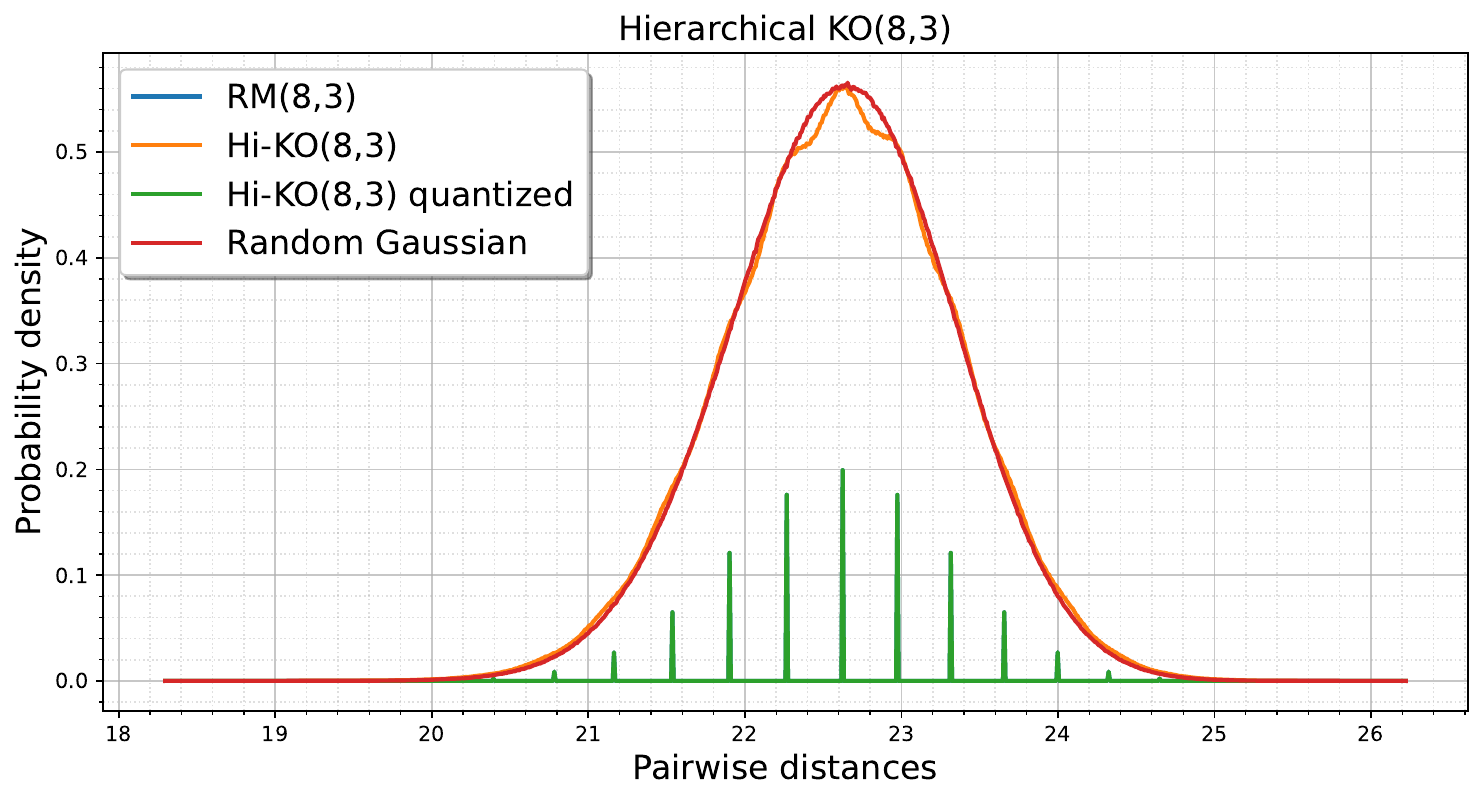}
	\caption{Pairwise distance distribution for RM(8,3), HiKO(8,3), quantized HiKO(8,3), and random Gaussian codebooks.}
	\label{fig:distance_distribution}
\end{figure}

To gain deeper insights into HiKO codes, we analyze the pairwise Euclidean distance distribution between codewords. Figure \ref{fig:distance_distribution} compares the distance distributions of RM(8,3), HiKO(8,3), its quantized version, and a random Gaussian codebook.
The analysis reveals a striking resemblance between HiKO(8,3) and random Gaussian codebooks. While RM(8,3) exhibits a concentrated distance profile characteristic of algebraic codes, HiKO(8,3) achieves a broader, more Gaussian-like distribution. Even the quantized version of HiKO(8,3) maintains this favorable property while ensuring binary transmission compatibility. This provides fundamental insight into the performance of HiKO codes, which combine efficient decodability with near-optimal distance properties of random codebooks.

\subsection{Number of Trainable Parameters}

The HiKO encoder's neural network $\tilde{g}_i$ contains $(2l+1)H + 2(H+1)H + (H+1)l + 1$ trainable parameters, where $l$ represents the input codeword length and $H$ denotes the hidden layer width. In comparison, the original KO encoder's $\tilde{g}_i$ network has $(2l+1)H + 2(H+1)H + (H+1)l$ parameters. The single additional parameter in HiKO comes from the learnable skip connection we introduced.
Despite architectural enhancements, the overall parameter count in HiKO codes remains nearly identical to the original KO codes. Our improvements focus primarily on the training methodology, incorporating dropout regularization and learnable skip connections without significantly increasing model complexity. This hierarchical approach efficiently transfers knowledge from smaller constituent codes to larger target codes, achieving superior error correction performance without incurring additional computational overhead during inference.

\section{Conclusion}
\label{sec:conclusion}

This paper presented HiKO codes, a hierarchical framework for training beyond-second-order Kronecker Operation neural codes. Our approach addresses the limitations of standard KO codes at higher rates through: (1) a hierarchical training methodology leveraging knowledge from constituent codes, (2) enhanced neural architectures with improved regularization, and (3) progressive unfreezing for stable optimization.
Experimental results demonstrated that HiKO codes consistently outperform classical RM codes across multiple parameter configurations, with HiKO(9,3) showing the most significant improvements at practical SNR values. Analysis of codeword distance distributions revealed that HiKO codes successfully approximate optimal Gaussian codebooks while maintaining efficient decoding properties.
The HiKO framework represents the first successful extension of neural codes to higher rates, opening new possibilities for practical deployment in high-throughput communication systems while maintaining the same parametric complexity as KO codes.


\end{document}